\documentclass{moriond}

\def\be{\begin{equation}}
\def\ee{\end{equation}}
\def\bea{\begin{eqnarray}}
\def\eea{\end{eqnarray}}

\def\babar{\mbox{\slshape B\kern-0.1em{\smaller A}\kern-0.1em B\kern-0.1em{\smaller A\kern-0.2em R}}}

\newcommand{\Photo}{{}}

\begin{document}
\vspace*{4cm}
\title{Status of the Belle II experiment before the first beams}

\author{ R. Cheaib}

\address{Department of Physics, University of Mississippi, P.O. Box 38677 \\
USA}
\maketitle

\abstracts{ The Belle II experiment has been preparing for its first $e+e-$ collisions, scheduled in April 2018. With a  target luminosity 40 times greater than the Belle experiment, the goal of Belle II  is to open the door to a panorama of measurements in heavy flavour physics and much more. The initial beams and the commissioning of the SuperKEKB collider took place earlier in 2016. After completing most of  the detector assembly and performing test runs with cosmic rays, the experiment is ready for initial collisions. Soon after in early 2019, data collection with the full Belle II detector will commence and the largest {\it B}-factory data sample will be collected. }

\section{Introduction}
The SuperKEKB collider and the Belle II detector~\cite{belle2} comprise the next generation $B$-meson factory, the Belle II experiment. Its predecessors  Belle~\cite{belle} and BaBar ~\cite{babar} have been very successful in performing precision measurements in heavy flavour physics using a total data sample with an integrated luminosity of about 424 and 711 fb$^{-1}$, respectively. Both Belle and BaBar confirmed many Standard Model (SM) predictions and this success lead to the awarding of the 2008 Nobel Prize in physics to M. Kobayashi and T. Maskawa~\cite{nobel}. 

The focus of the Belle II experiment is to push the boundaries of new physics searches at the intensity frontier in the relatively clean environment of $e+e-$ collisions at the GeV scale. The plan for the Belle II experiment is to collect a data sample with a total integrated luminosity of  50 ab$^{-1}$ over the period of 7 years. The high statistics sample will allow for many interesting measurements in the heavy flavour and lepton sector. Furthermore, many new physics searches and anomalies will be unraveled and probed. To achieve this goal, the target luminosity of the SuperKEKB collider will be 8.5 x 10$^{35}$ s$^{-1}$$cm^{-2}$, which is 40 times higher than that of KEKB. Furthermore, the Belle II detector must be upgraded such that it can accommodate the anticipated increase in luminosity and control limiting factors such as the high levels of beam-related backgrounds.

\section{SuperKEKB and the Belle II detector}\label{}

The SuperKEKB collider is the extension of KEKB to higher luminosities, based on the ``nano-beam” scheme proposed by P. Raimondi from Frascati ~\cite{frascati}. In this scheme, the vertical size of the beam is ``squished`` such that vertical beta function $\beta_y$ at the interaction point (IP) is decreased by a factor of 20. This is done by minimizing the longitudinal size of the overlap region of the two beams and results in a factor of 20 improvement in the luminosity. Furthermore, the intensity of the beam currents is also doubled  to achieve the luminosity goal of SuperKEKB.

The Belle II detector has been upgraded to ensure its performance will be maintained with the higher background and physics event rates that accompany the increase in luminosity. Inheriting from the Belle detector design~\cite{belle}, the inner part of the Belle II detector consists of two layers of high granularity pixel sensors (VXD) in the inner side and four layers of double-sided silicon strip detectors (SVD)  in the outer radii. The Belle II central drift chamber (CDC) surrounds the SVD and has smaller cells as well as fast readout electronics to provide improved measurements of $dE/dx$, momenta, and particle trajectories. A new time-of-propagation (TOP) Cherenkov counter is installed in the barrel region and can yield better separation between kaons and pions.  Another ring-imaging Cherenkov detector with aerogel radiators (A-RICH) is also placed in the end-caps to maintain the charged particle identification performance. The electromagnetic calorimeter uses the same scintillator crystals of the Belle detector, while providing an improved signal-to-background separation due to waveform sampling. Finally, the resistive plate chambers in the end-cap region of the K$_{L}$ and muon detector are replaced with scintillators to tolerate the high background rates in the Belle II environment. Overall, the Belle II detector has better performance than Belle, despite the higher background rates and harsher environment. 

\section{Phase I: Commissioning and the BEAST detector }
The first phase of the commissioning of the Belle II detector was completed in June 2016. During phase I, high positron and electron currents were circulated with the SuperKEKB collider without any collisions. The goal of the first phase was to clean the beam pipe, tune the accelerator optics and monitor the beam conditions in real time. The latter was done using a  system of radiation detectors, referred to as BEAST II,  which are placed on the beam line and are used to collect beam background data in the detector.  The main sources of beam backgrounds, in the absence of collisions, are Touschek scattering and beam-gas scattering. The former results from Coulomb scattering between two particles in the same bunch, while the latter is due to scattering off residual gas atoms in the the beam pipe. A full report on the results of the measurements with the BEAST II detectors will be published soon~\cite{beast}.

After collecting 5 months of data with the BEAST II detectors,  phase I was complete. Shortly after, the TOP was installed in May 2016, followed by the drift chamber in October and the A-RICH in  August, 2017. The final focusing magnets were also installed around the same time.  Finally, in April 2017, the Belle II detector was rolled in to the beam line. Global cosmic runs took place with the 1.5 T magnetic field throughout July and August 2017 and tested the integration of all the installed sub-detectors as well as the data acquisition system.
\section{Phase II: Initial Beams}

\begin{figure}
\centerline{\includegraphics[width=8cm]{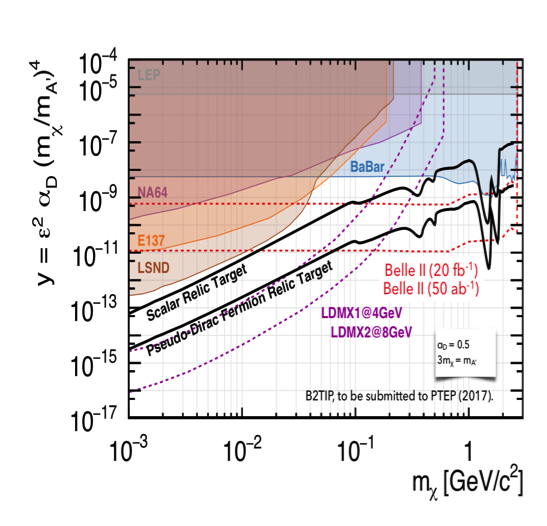}}
\caption{ Predicted sensitivity of the Belle II data samples for a given light dark matter mass $m_{\chi}$ as a function of the mixing parameter $\epsilon$ and the mass of the dark photon $m_{A'}$.}
\label{fig:Fig1}
\end{figure}

The plan for phase II is to run collisions with the final focusing magnets installed and the detector in place. At this point, however, the VXD is not yet installed. Phase II recently started and the plan is to collect a data sample with a total integrated luminosity of 20 fb$^{-1}$. 
Without the VXD, phase II studies should rely mainly on tracking from the CDC, as well as calorimeter-reliant measurements from the ECL.  The list of early physics topics include:  Bottomonium-like searches  below and above the $\Upsilon(4S)$ , dark sector and light Higgs searches, and tau lepton decays. 
Various topics of quarkonia can be examined with a relatively small dataset at an energy other than the $\Upsilon(4S)$. For instance, there are four predicted Bottomonium states that are  yet to be identified below the  $\Upsilon(3S)$, such as $\eta_{b}(3S)$. Furthermore, many important parameters related to known states are yet to be measured with accuracy and Belle II is expected to have improved sensitivity to the related channels. 
In addition, interesting searches in the dark sector can be done with the phase II dataset. For example, the search for a dark photon in the decay $e^{+}e^{-}\rightarrow \gamma A'$, where A' decays invisibly into a pair of light dark matter candidates, can be done using Phase II data. The search requires a trigger to record events in which the final state consists of only a single photon.  Because the  phase II luminosity is low, the triggers could be configured to be looser, thus allowing for an interesting measurement. With the improved performance of the Belle II detector, the sensitivity to many of the dark sector channels will be competitive, as shown for the case of the dark photon search in Fig~\ref{fig:Fig1}.

\section{Phase III}
At the beginning of 2019, the first full physics run of the Belle II experiment is expected. Collisions at the nominal energy will take place with the full detector installed, including the VXD. With a target integrated luminosity of 50 ab$^{-1}$, the physics agenda for Belle II is vast and covers a wide range of topics. This includes {\it CP} violation studies, rare {\it B}  decays, lepton flavour violation, charm physics, D-mixing, and much more.
One of the highly anticipated Belle II results on the radar is the measurement of R(D$^{(*)}$).  R(D$^{(*)}$) is  given by the equation below:

\begin{equation}
R(D^{(*)}) = \frac{\cal B(B\rightarrow D^{(*)}\tau\nu_{\tau} )}{\cal B(B\rightarrow D^{(*)} \ell \nu_{\ell} )}
\end{equation}

As shown in Fig 2, previous measurements by BaBar, Belle, and LHCb reveal a combined 4.1 $\sigma$ deviation from the SM expectation\cite{dtaunu}.  

\begin{figure}
\centerline{\includegraphics[height=6cm]{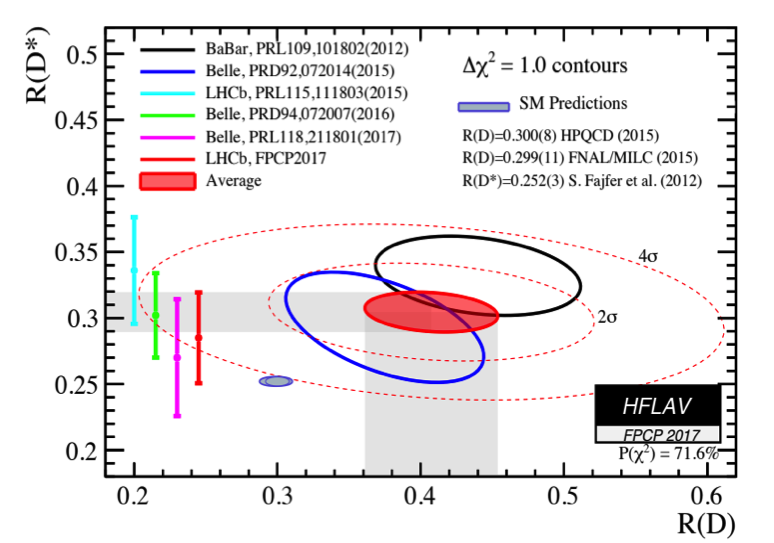}}
\label{fig:FigS}
\caption{Recent measurements of R(D) and R(D*) by Belle, BaBar, and LHCb.}
\end{figure}

\begin{figure}
\centerline{\includegraphics[height=3cm]{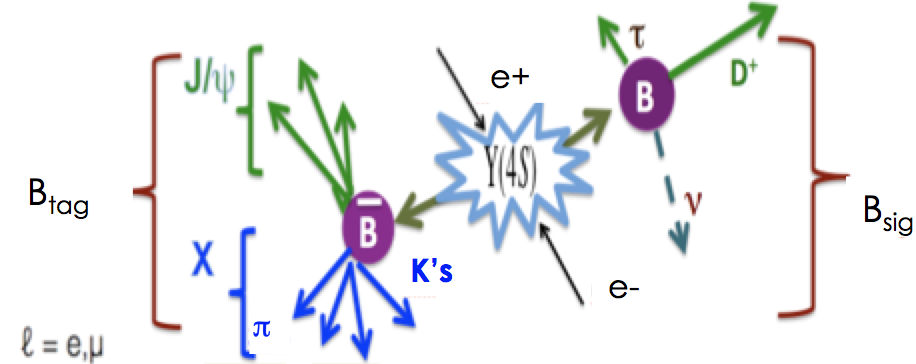}}
\caption{ Schematic of the Full Event Interpretation approach.}
\label{fig:Fig3}
\end{figure}

The goal of the Belle II experiment is to unravel this B-anomaly by pushing the significance of the R(D) and R(D*) measurements to the boundaries of discovery. Belle II is planning to use a novel technique, called the Full Event Interpretation, where one B meson, referred to as $B_{tag}$ can be reconstructed exclusively via hadronic modes. The search for the signal, $ B_{sig} \rightarrow D^{(*)}\tau\nu_{\tau}$ is then done using information in the rest of the event, as shown in Fig \ref{fig:Fig3}. The FEI approach is ideal for decays with neutrinos, since any missing energy in the event will be attributed to $B_{sig}$. Furthermore, with the novel FEI,  it is possible to first search for the signal decay before reconstructing the $B_{tag}$ in the event.  This implies using a FEI approach that is optimized to a specific signal decay and thus increase the signal efficiency. With this approach, observation of new physics, if it exists, can be expected after collecting about 5$ab^{-1}$ of data, only two years after the start of Phase III.

\section{Summary}
The Belle II experiment has recently started phase II with the detector in place, except for the VXD. A data sample of 20fb$^{-1}$ will be collected, possibly at energies other than the $\Upsilon(4S)$, and a comprehensive early physics program has been developed. Soon after, the Belle II experiment will start collecting data with the full detector intact and phase III will start. A wide range of physics topics will be covered and the upcoming era will be filled with anticipation and excitement as many new results will potentially unfold. Stay tuned!

\end{document}